\documentstyle[aps,twocolumn,pra,graphicx]{revtex}

\newcommand{\ket}[1]{{\left|#1\right\rangle}} 
\newcommand{\bra}[1]{{\left\langle#1\right|}} 
\newcommand{\kvec}{{\mathbf k}} 
\newcommand{\rvec}{{\mathbf r}} 
\newcommand{\Rvec}{{\mathbf R}} 
\newcommand{\dg}{^\dagger} 

\newcommand{\Tr}{{\text{Tr}}}

\begin{document}                
\draft
\wideabs{
\author{T. Aichele, A. I. Lvovsky\footnote{Email: Alex.Lvovsky@uni-konstanz.de}}
\address{Fachbereich Physik, Universit\"at Konstanz, D-78457 Konstanz, Germany}
\author{S. Schiller}
\address{ Institut f\"ur
Experimentalphysik, Heinrich-Heine-Universit\"at D\"usseldorf,
D-40225 D\"usseldorf, Germany}
\date{\today}

\title{Optical mode characterization of single photons prepared via
conditional measurements on a biphoton state}
\maketitle

\begin{abstract}
A detailed theoretical analysis of the spatiotemporal mode of a
single photon prepared via conditional measurements on a photon pair
generated in the process of parametric down-conversion is presented.
The maximum efficiency of coupling the photon into a
transform-limited classical optical mode is calculated and ways for
its optimization are determined. An experimentally feasible technique
of generating the optimally matching classical mode is proposed. The
theory is applied to a recent experiment on pulsed homodyne
tomography of the single-photon Fock state (A. I. Lvovsky et al.,
Phys Rev. Lett. {\bf 87}, 050402 (2001)).
\end{abstract}

\pacs{} }PACS numbers: 42.50.Dv Nonclassical field states; squeezed,
antibunched, and sub-Poissonian states; operational definitions of
the phase of the field; phase measurements; 42.50.Ar Photon
statistics and coherence theory; 03.65.Ud Entanglement and quantum
nonlocality

\section{Introduction.} Quantum states containing a definite number of
energy quanta (Fock states) play a key role in quantum optics. They
constitute the essence of the quantum nature of light and are
indispensable in the theoretical description of a wide range of
optical phenomena. Fock states also play a major role in more applied
aspects of quantum optics, such as the rapidly developing fields of
quantum communication and information. Of particular importance is
the application of Fock states in quantum cryptography which would
result in a significant increase in capacity and security of
communication channels \cite{Gisin}. Superpositions of the vacuum and
the single-photon state in a certain optical mode can also be used to
implement a qubit. Such an application of the Fock state has been
discussed in a recent proposal on efficient linear quantum
computation \cite{KLM}.

Despite their importance, pure number states are extremely rare in
nature
and their synthesis in a laboratory constitutes
a rather involved task.
In recent years significant efforts have been made towards developing
a ``photon pistol" --- a technology of emitting a single photon into
a well-defined traveling spatiotemporal mode upon the onset of a
classical trigger. A number of approaches are being tried
\cite{pistol}, but none has fully resolved this problem so far. In
these circumstances an important alternative is offered by the
technique of preparing single photons by conditional measurements on
a biphoton state born in the process of parametric down-conversion
(PDC) (Fig.\,1). In PDC, a ``pump" photon propagating through a
nonlinear medium may spontaneously annihilate to produce two photons
of lower energy in the form of a highly entangled quantum state known
as $biphoton$. The two generated photons are separated into two
emission channels according to their propagation direction,
wavelength and/or polarization. Detection of a photon in one of the
emission channels (labeled $trigger$) causes the non-local photon
pair to collapse projecting the quantum state in the remaining
($signal$) channel into a single-photon state. Proposed and tested
experimentally in 1986 by Hong and Mandel \cite{HM} as well as
Grangier, Roger and Aspect \cite{GRA}, this technique has become a
workhorse for many quantum optics experiments \cite{workhorse}.

\begin{figure}
\begin{center}
\includegraphics[width=0.35\textwidth]{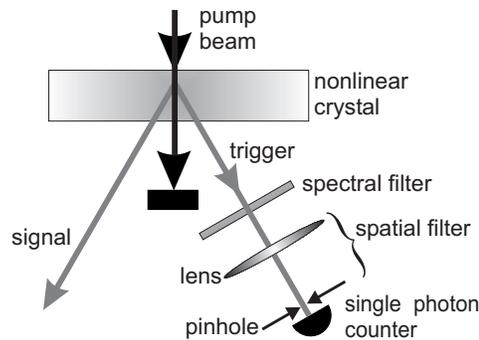}
\caption{\label{AbbPDC1}Preparation of single photons by conditional
measurements on a biphoton state.}
\end{center}
\end{figure}

Using the conditionally prepared photon (CPP) for practical purposes,
such as communication, storage, quantum information processing or
synthesis of more complex quantum states, requires the photon to be
produced in a well-defined, transform-limited optical mode. While
there exists extensive theoretical and experimental research on
biphotons and associated quantum effects
\cite{corr1,RubinStrekalov,corr2,Walmsley}, very little work has been
done to analyze the CPP as a ``final product", i.e. in a way that
would reveal its quantum state and provide ways for its optimization.

In 1997 Ou presented a qualitative theoretical investigation of the
temporal mode of the CPP in application to homodyne tomography
\cite{Ou}. He has shown that in order to prepare the photon in a
transform limited temporal mode that could be matched to the local
oscillator beam, one has to use a spectral filter in the trigger
channel which is much narrower than the linewidth of the pump pulse.
In a more recent paper, Grosshans and Grangier considered a similar
experiment and quantified the fidelity of mode matching in terms of
the efficiency of the tomography measurement \cite{Grangier}. This
efficiency was calculated using specific assumptions about the local
oscillator mode and a simplified model of the spectral filter has
been used. Very recently, the Fock state tomography has been
demonstrated experimentally \cite{Fock}, but no theoretical
discussion regarding the above matters has been presented.

In this paper we perform a detailed theoretical analysis of the
spatiotemporal mode of the CPP using very general assumptions about
the spatial and spectral filter in the trigger channel. We calculate
the distribution of the CPP state over plane wave modes and discuss
ways of its optimal mode matching to a transform-limited classical
wave. We determine the theoretical limits imposed on the mode
matching parameter and propose a method of constructing a classical
field that would match the CPP mode optimally.

\section{The conditionally prepared single photon}
We start with a general calculation of the quantum state of a single
photon state prepared by a conditional measurement on a pulsed PDC
biphoton state. We restrict our consideration to the pulsed regime. A
continuous-wave pump, although highly efficient in generating
biphotons \cite{Weinfurter}, does not yield transform-limited CPPs
 as we demonstrate below. In all calculations in this paper, we neglect polarization
entanglement (polarization is assumed to be well-defined in both PDC
channels) and refraction inside the crystal.

The interaction Hamiltonian of parametric down-conversion is given by
\cite{OHM}
\begin{eqnarray}
  \hat V&&(t) \\&&= \alpha \int\tilde K(\rvec) \tilde E_t^{(-)}(\rvec,t)
  \tilde
  E_s^{(-)}(\rvec,t) \tilde E_p^{(+)}(\rvec,t) d^3r \, + \text{H.c.}
  ,\nonumber
\end{eqnarray}
where $\alpha$ is proportional to the second order nonlinear
susceptibility and is assumed frequency independent, $\tilde
K(\rvec)$ describes the nonlinear crystal volume and is one inside
and zero outside the crystal. We treat the fields in the signal ($s$)
and trigger ($t$) channels as quantum operators, with their
positive-frequency components given by
\begin{equation}
  \hat{\tilde E}_{s,t}^{(+)}(\rvec,t)\label{E_s} =\int
  e^{-i(\kvec_{s,t}\cdot\rvec-\omega_{s,t} t)}
  \hat a_{\kvec_{s,t},\omega_{s,t}} d^3k_{s,t} \, d\omega_{s,t};
\end{equation}
the coherent pump field is treated classically:
\begin{equation}
  \tilde E_p^{(+)}(\rvec,t) = \int E_p^{(+)}(\kvec_p,\omega_p)
  e^{i(\kvec_p\cdot\rvec-\omega_p t)} d^3k_p \, d\omega_p \, .
\end{equation}
For all fields, quantum or classical, the Hermitian electric field
observable is written as $\tilde E_p(\rvec,t)=\tilde
E_p^{(+)}(\rvec,t)+\tilde E_p^{(-)}(\rvec,t)$, with $\tilde
E_p^{(-)}(\rvec,t)=(\tilde E_p^{(+)}(\rvec,t))\dg$.

Assuming the signal and trigger modes to be initially in the vacuum
state and restricting the consideration to the first order
perturbation theory, we write the resulting biphoton state as
\begin{equation}
\ket B = \ket{0}_s\ket{0}_t - i \int_{-\infty}^{\infty}
 \hat V(t)\,dt .
\end{equation}
Performing the integration we obtain:
\begin{eqnarray}
  \ket B &=&  \ket{0}_s\ket{0}_t - i \int d^3k_s \, d\omega_s \, d^3k_t \,
  d\omega_t \nonumber\\
  && \times \, \Psi(\kvec_s,\omega_s,\kvec_t,\omega_t)
  \ket{1_{\kvec_s,\omega_s}}_s \ket{1_{\kvec_t,\omega_t}}_t,
\end{eqnarray}
with
\begin{eqnarray}
  \Psi(\kvec_s,\omega_s,&\kvec_t&,\omega_t)\nonumber\\ &=&
  \alpha\int E_p^{(+)}(\kvec_p,\omega_s+\omega_t) K(\Delta\kvec)
  d^3k_p .
\label{EqnPsi}
\end{eqnarray}
Here $K(\kvec)$ is the Fourier transform of $\tilde K(\rvec)$ and the
$\kvec$-vector mismatch is $\Delta\kvec=\kvec_p-\kvec_s-\kvec_t$.


The trigger photon is then selected by spatial and frequency filters
and is detected by a single-photon counter. Conditioned on the
detection event the non-local biphoton state collapses into a single
photon state in the signal mode. The properties of this mode are
determined by the optical mode of the pump photon and the spatial and
spectral filtering in the trigger channel:
\begin{equation}
  \hat\rho_s=\text{Tr}_t(\hat \rho_t\ket{B}\bra{B})
  \label{rho_s_Tr},
\end{equation}
where the trace is taken over the trigger states and $\hat\rho_t$
denotes the state ensemble selected by the filters:
\begin{equation}
  \hat{\rho}_t = \int
  T(\kvec_t,\omega_t) \ket{1_{\kvec_t,\omega_t}}_t \bra{1_{\kvec_t,\omega_t}}_t \,
  d^3k_t \, d\omega_t \, \label{rho_t}
\end{equation}
with $T(\kvec,\omega)$ being the spatiotemporal transmission function
of the filters.

The expression (\ref{rho_t}) is different from the one used by
Grosshans and Grangier \cite{Grangier} who associated a monochromator
of width $\delta\omega$ with a pure state of the form $\ket{\psi_t}=
\int_{-\delta\omega/2}^{\delta\omega/2}\ket{1_{\omega_0+\delta\omega}}\,d\delta\omega$.
A spectral filter does not distinguish relative phases of different
frequency components of the transmitted ensemble;
we have therefore assumed all non-diagonal elements of the trigger
density matrix to vanish.

An explicit calculation of the quantum state (\ref{rho_s_Tr}) of the
photon in the signal channel yields
\begin{eqnarray}        \label{rho_s}
  \hat\rho_s=&& \int d^3k_s d\omega_s d^3k_s'  d\omega_s'  \\ &&
  \Phi(\kvec_s,\omega_s,\kvec_s',\omega_s') \nonumber
   \ket{1_{\kvec_s',\omega_s'}}_s
  \bra{1_{\kvec_s,\omega_s}}_s ,
\end{eqnarray}
where
\begin{eqnarray}
  \Phi(\kvec_s,\omega_s,\kvec_s',\omega_s') &=& |\alpha|^2 \int d^3k_t \,
  d\omega_t \, d^3k_p \, \, d^3k_p'\, \nonumber\\
  && \times \, E_p^{(-)}(\kvec_p,\omega_s+\omega_t)
  E_p^{(+)}(\kvec_p',\omega_s'+\omega_t) \nonumber\\
  && \times \, T(\kvec_t,\omega_t) K^*(\Delta\kvec)
  K(\Delta\kvec'),
\label{EqnQMech}
\end{eqnarray}
with $\Delta\kvec$ as above and $\Delta\kvec'=\kvec_p'-\kvec_s'-\kvec_t$.

\section{The measure of mode matching}
To discuss the main question of this paper --- how to match a
classical wave to the spatiotemporal mode of the CPP
--- we first need to introduce a quantitative measure of mode
matching. We characterize both modes by their correlation functions,
defined as $\Gamma(\kvec,\omega,\kvec',\omega')={<\hat
E{(-)}(\kvec,\omega)\hat E{(+)}(\kvec',\omega')>}$, with the
averaging done in the statistical sense for the classical field and
in the quantum-mechanical sense for the single-photon field. For the
latter, using $\hat E^{(+)}(\kvec,\omega)\propto\hat
a_{\kvec,\omega}$ and applying Eq. (\ref{rho_s}), we find:
\begin{equation}\label{GammaPhi}
\Gamma (\kvec,\omega,\kvec',\omega')={\rm Tr}(\hat\rho_s\,\hat
a^\dagger_{\kvec,\omega}\,\hat
a_{\kvec',\omega'})=\Phi(\kvec,\omega,\kvec',\omega'),
\end{equation}
i.e. the field correlation function coincides with the density matrix
of the single photon state.

It is natural to define the degree of mode matching between two waves
characterized by their correlation functions $
\Gamma_{1,2}(\kvec,\omega,\kvec',\omega')$ as follows:

\begin{equation}\label{EqnDefMM}
M = \frac{\int d^3k \, d\omega \,  d^3k' \, d\omega' \,
  \Gamma_1(\kvec,\omega,\kvec',\omega')
  \Gamma_2^*(\kvec,\omega,\kvec',\omega')}{\int d^3k \, d\omega \,
  \Gamma_1(\kvec,\omega,\kvec,\omega) \int d^3k \, d\omega \,
  \Gamma_2(\kvec,\omega,\kvec,\omega)}.
\end{equation}
If both waves $\Gamma_1$ and $\Gamma_2$ are classical, the mode
matching parameter is equal to the square of the visibility of the
pattern that would be observed if the two modes were caused to
interfere. If both waves are single photons, the value of $M$ is the
probability of the quantum overlap ${\rm Tr}(\hat\rho_1\hat\rho_2)$
between the two states. We are most interested in the third case,
when one of the $\Gamma$'s represents a CPP, and the other a matching
classical wave, and adopt the above expression as the measure of mode
matching. Grosshans and Grangier \cite{Grangier} have shown that the
expression (\ref{EqnDefMM}) determines the quantum efficiency in a
homodyne tomography measurement of the single-photon Fock state in
which the matching classical wave serves as a local oscillator.

Suppose that a single photon is prepared in a certain state
$\hat\rho_s$ and our task is to pick the classical wave that would
match the mode of the single photon optimally. As the former is
generally not a pure quantum state, no choice of the classical mode
can guarantee perfect mode matching. To determine the maximum level
of $M$ that can be achieved, we introduce the {\it purity parameter}
of an optical mode,
\begin{equation} \label{P_s}
P=\frac{\int d^3k \, d\omega \,  d^3k' \, d\omega' \,
  \Gamma(\kvec,\omega,\kvec',\omega')  \Gamma^*(\kvec,\omega,\kvec',\omega')}
  {(\int d^3k \, d\omega \,
  \Gamma(\kvec,\omega,\kvec,\omega))^2},\nonumber
\end{equation}
which is equal to unity for coherent optical modes and vanishes for
incoherent ones. For the single-photon states of the form
(\ref{rho_s}), the above quantity can be written in the form of a
well-known quantum state purity parameter
\begin{equation}
P=\Tr(\hat\rho_s^2),
\end{equation}
which reaches one for pure quantum states and approaches zero for
density matrices with no non-diagonal elements.

It then follows from the Cauchy-Schwartz inequality that for any two
optical modes 1 and 2
\begin{equation}
M^2\le P_1\,P_2.
\end{equation}
If mode 1 is a CPP, the right-hand side of the above inequality is
maximized if the matching classical mode 2 is a coherent wave,
i.e. $P_1=1$. In this case,
\begin{equation}  \label{limit}
M\le \sqrt{P_1},
\end{equation}
which establishes an unconditional theoretical limit to the degree to
which a classical wave can be matched to a given single-photon mode
(\ref{rho_s}).

\section{Modeling the single photon mode with a classical wave}  \label{SecModel}
Our next task is to design a classical wave that matches the CPP mode
optimally. Apart from its theoretical aspect, this problem
constitutes a substantial challenge in the experimental practice. The
traditional procedure of matching two classical modes with each other
--- by observing interference fringes and optimizing their visibility
--- is clearly not applicable to the situation when one of the modes
is a single photon. There is no laser beam to mode match to. The only
information available to the experimentalist is the remote location
and width of the trigger filter and the parameters of the pump.
Although the spatial location of the CPP can be approximately
determined by detecting coincidences between the photon count events
in the signal and trigger \cite{RubinStrekalov}, optimizing the mode
matching requires adjustment of a much larger set of degrees of
freedom, such as the beam direction, divergence, spatial and temporal
width, optical delay, etc. Reliable adjustment of these parameters
cannot be achieved through a sole optimization of the coincidence
rate.

\begin{figure}
\begin{center}
\includegraphics[width=0.45\textwidth]{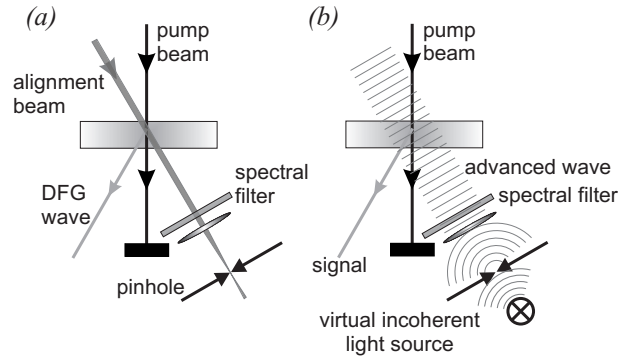}
\caption{\label{PDC2} (a) Nonlinear interaction of the alignment beam
with the pump pulse generates DFG emission; (b) Interaction of
Klyshko's advanced wave with the pump generates a DFG mode that
mimics that of the CPP.}
\end{center}
\end{figure}

Fortunately, the CPP mode can be modeled with a classical wave
generated in the following way. Suppose an {\it alignment beam} is
inserted into the trigger channel so that it overlaps spatially and
temporally with the pump beam inside the crystal and passes through
the optical filters (Fig. \ref{PDC2} (a)). Nonlinear interaction of
such an alignment beam with the pump wave will produce difference
frequency generation (DFG) into a spatiotemporal mode similar to that
of the CPP.

To show this, we write for the nonlinear polarization inside the
crystal
\begin{equation}\label{Eproduct}
  \tilde P_{\rm DFG}(\rvec,t) \propto  \tilde E_A(\rvec,t) \tilde E_p(\rvec,t).
\end{equation}
Here $\tilde E_p(\rvec,t)$ and $\tilde E_A(\rvec,t)$ are the electric
fields of the pump and alignment beams, respectively. The nonlinear
polarization gives rise to the DFG field which is obtained from Eq.
(\ref{Eproduct}) via a Fourier transform which is restricted to the
crystal volume:
\begin{eqnarray}
  E^{(+)}_{\rm DFG}&&(\kvec_s,\omega_s) = \beta' \,\delta(k_s-\omega_s/c)\int d^3k_A \, d\omega_A \, d^3k_p \nonumber
  \\ \times&&
  E_A^{(-)}(\kvec_A,\omega_A) \, E_p^{(+)}(\kvec_p,\omega_s+\omega_A) K(\Delta\kvec).
\end{eqnarray}
The proportionality coefficient $\beta$ represents the nonlinearity
of the medium.
 If the alignment field
is partially incoherent and is characterized by a correlation
function $ \Gamma_A(\kvec_A,\omega_A,\kvec_A',\omega_A')$, the above
equation generalizes to
\begin{eqnarray}  \label{EqnKlass}
  \Gamma_{\rm DFG}(\kvec_s,\omega_s,&&\kvec_s',\omega_s')
   = |\beta'|^2 \,\delta(k_s-\omega_s/c)\,\delta(k'_s-\omega'_s/c)\\&&\times\int d^3k_A \,
  d\omega_A  \, d^3k_A' \, d\omega_A' \, d^3k_p \, d^3k_p' \nonumber\\
  && \times \, E_p^{(-)}(\kvec_p,\omega_s+\omega_A)
  E_p^{(+)}(\kvec_p',\omega_s'+\omega_A') \nonumber\\
  && \times \, \Gamma_A^*(\kvec_A,\omega_A,\kvec_A',\omega_A') K^*(\Delta\kvec)
  K(\Delta\kvec'). \nonumber
\end{eqnarray}

We immediately notice that the expressions for the optical mode of
the CPP photon (\ref{EqnQMech}) and of the DFG pulse (\ref{EqnKlass})
are very similar\footnote{The delta-functions, included into Eq.
(\ref{EqnKlass}) to eliminate nonphysical Fourier components of the
DFG field, are also implicitly present in Eqs. (\ref{rho_s}) and
(\ref{EqnQMech}) as the single-photon states
$\ket{1_{\kvec_s,\omega_s}}_s$ exist only when $k_s=\omega_s/c$.}.
This similarity can be interpreted in the framework of D. N.
Klyshko's concept of advanced waves \cite{Klyshko}. Suppose the
single photon detector is replaced by an incoherent source (an
``incandescent bulb") continuously emitting omnidirectional
incoherent light into a wide spectral range backwards in time. This
completely incoherent light is characterized by the correlation
function $\Gamma_0(\kvec',\omega',\kvec,\omega)=
\delta^{(3)}(\kvec'-\kvec)\,\delta(\omega'-\omega)$ which, upon
passing through the spatial and spectral filters, transforms into
\begin{equation}  \label{filter}
\Gamma_t(\kvec',\omega',\kvec,\omega)=T(\kvec,\omega)\,
\delta^{(3)}(\kvec'-\kvec)\,\delta(\omega'-\omega).
\end{equation}
The advanced wave then enters the nonlinear crystal and interacts
with the pump wave whenever and wherever it is present in the
crystal. The nonlinear interaction of Klyshko's advanced wave with
the pump pulse produces a pulse of DFG emission into the signal
channel (Fig.\,\ref{PDC2}(b)). Substituting the correlation function
(\ref{filter}) of the advanced wave into Eq.\,(\ref{EqnKlass}) as
$\Gamma_A$ we find that {\it the correlation function $\Gamma_{\rm
DFG}(\kvec_s,\omega_s,\kvec_s',\omega_s')$ of the DFG pulse generated
through the nonlinear interaction of the advanced wave and the pump
pulse is identical to the density matrix
$\Phi(\kvec_s,\omega_s,\kvec_s',\omega_s')$ of the single photon
prepared via conditional measurements on a biphoton performed in the
same optical arrangement.}

This identity can be easily generalized to optical filters of random
configuration, more complex than a combination of spatial and
spectral filters described by Eqs. (\ref{rho_t}) and (\ref{filter}).
Its applicability is also independent from other features of the
experimental setup, such as the type of PDC, properties of the pump
beam, geometry of the crystal, walk-off and group velocity dispersion
effects, {\it etc.} and appears to be very general. The only
restriction that has to be taken into account is the first order
perturbation theory that implies that the probability of generating
two or more biphotons at a time is negligible.

By varying the configuration of the filter in the trigger channel
one has a degree of freedom in forming the CPP mode with required
spatiotemporal properties. This possibility can be considered as
an example of {\it remote state preparation} in the sense
discussed by Bennett {\it et al.} \cite{RSP}. The original
biphoton state is highly entangled in the frequency-momentum space
and this entanglement plays an essential role in generating the
Fock state. The signal mode {\it does not exist} unless and until
the trigger photon passes through the filters and is registered. A
detection event results in a non-local preparation of a single
photon in an optical mode whose characteristics are determined by
the way in which the measurement in the trigger channel is
performed.

\section{Mode matching: an explicit calculation} \label{SecCalc}
\subsection{Frequency-momentum representation}

We have thus designed an experimentally plausible way of generating a
classical wave whose mode models that of the CPP. Although the
advanced (i.e. propagating backwards in time) wave is a purely
imaginary object, it can be simulated in a laboratory by a coherent
laser beam --- the alignment wave. As we demonstrate in this section,
the proper choice of the latter allows the level of mode matching to
reach its theoretical limit set by Eq. (\ref{limit}). To simplify our
calculations, we make the following assumptions.


1. Parametric down-conversion occurs in a collinear type II
configuration. The signal and trigger channels are then separated
according to their polarization. Collinearity of the pump, signal and
trigger fields allows to use the same reference frame for all three
waves.

2. A simple combination of spatial and spectral filters is used in
the trigger channel, so Eqs. (\ref{rho_t}) and (\ref{filter}) are
valid.

3. The crystal volume is much larger than the spatial extent of the
pump pulse inside the crystal. This allows us to approximate
\begin{equation} \label{largecrys}
V(\Delta\kvec)\approx\delta^{(3)}(\Delta\kvec)
\end{equation} and Eqs.\,(\ref{EqnQMech}) and (\ref{EqnKlass})
simplify accordingly.

4. The pump ($p$) and alignment ($A$) fields are collimated inside
the crystal and are assumed to be of Gaussian shape:
\begin{equation} \label{EpA_omega}
  E_{p,A}^{(+)}(\kvec,\omega) = E_{p,A}^0
  \exp\left(-\frac{\kvec_{\bot}^2}{\kappa_{p,A}^2}
  -\frac{(\omega-\omega_{p,A}^0)^2}{\sigma_{p,A}^2}\right),
\end{equation}
where $\omega_{p,A}^0$ are the central frequencies of the two waves,
$\sigma_{p,A}$ are their linewidths and $\kappa_{p,A}$ are the beam
widths in the momentum space. A similar assumption is made for the
transmission of the trigger filter:
\begin{equation} \label{T_omega}
  T(\kvec,\omega) = T_0
  \exp\left(-\frac{\kvec_{\bot}^2}{\kappa_{t}^2}
  -\frac{(\omega-\omega_{t}^0)^2}{\sigma_t^2}\right).
\end{equation}

5. The center frequency of the alignment field coincides with the
transmission maximum of the spectral filter, i.e.
$\omega_A^0=\omega_t^0$ and its direction is collinear with the
transmission maximum of the spatial filter.

6. There is no beam walkoff nor group velocity dispersion.

Under these assumptions, the calculation of the mode matching and the
purity parameter substantially simplifies as the temporal variables
separate from the two spatial ones and the resulting values of $M$
and $P$ are products of three similar expressions for the temporal
and spatial (in two orthogonal dimensions) mode matching and mode
purity parameters. In this subsection, we restrict ourselves to the
temporal domain keeping in mind that the calculation in the spatial
domain would be completely analogous.

The density matrix of the CPP mode are determined using Eq.
(\ref{EqnQMech}):
\begin{eqnarray}\label{EqnGammaInc}
  &&\Phi(\omega,\omega')\\&& \,=  \Gamma_s^0
  \exp\biggl(-\frac{(\omega-\omega_s^0)^2 +
  (\omega'-\omega_s^0)^2}{(\sigma_p^2 + 2\sigma_t^2)}
  -\frac{\sigma_t^2(\omega-\omega')^2}{\sigma_p^2(\sigma_p^2 + 2\sigma_t^2)}\biggl). \nonumber
\end{eqnarray}
where $\Gamma_0^s$ is a constant factor and
$\omega_s^0=\omega_p^0-\omega_t^0$. An application of Eq. (\ref{P_s})
to the above expression yields the temporal purity parameter of the
CPP mode:
\begin{equation}  \label{P_temp}
P_{\rm temp}(\mu_t)=1/\sqrt{1+2\mu_t^2},
\end{equation}
with   $\mu_{t} = \sigma_{t}/\sigma_p$.

\begin{figure}
\begin{center}
\includegraphics[width=0.45\textwidth]{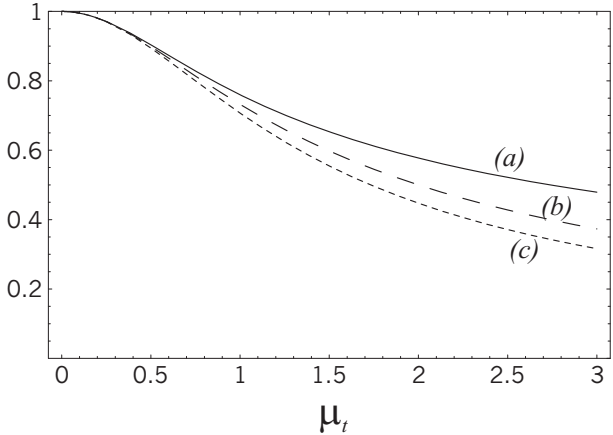}
\caption{\label{AbbMM} (a) Square root of the CPP state purity
parameter $\sqrt{P_{\rm temp}(\mu_t)}$ which sets the maximum
achievable level of mode matching for a given optical arrangement;
(b) temporal mode overlap $M(\mu_{t},\mu_{A}^{\rm max})$ between the
DFG and CPP modes for the optimally chosen alignment beam; (c)
temporal mode overlap $M(\mu_{t},0)$ for a plane wave alignment beam.
}
\end{center}
\end{figure}

The expression (\ref{P_temp}) (Fig. \ref{AbbMM}(a)) confirms the
conclusion of Ou \cite{Ou}: {\it narrowband filtering in the trigger
channel is crucial in obtaining a CPP mode that approaches a pure
state}. Only in this case can one achieve efficient mode matching
between the CPP and a classical mode. In this aspect our approach is
very different from the one taken by Grosshans and Grangier
\cite{Grangier}. According to their model, the ensemble selected by
the filter in the trigger channel is a pure state; as a consequence,
one can always achieve a perfect mode matching fidelity by picking
the proper parameters of the matching classical wave. As demonstrated
above, using the density matrix formalism to model the state ensemble
selected by the trigger leads to an $intrinsic$ reduction of the CPP
mode purity that cannot be compensated by adjusting the properties of
the matching classical wave.

This result contrasts with some recent reports demonstrating that
high-visibility quantum interference effects can be observed without
any  spectral filtering in the down-conversion channels
\cite{corr2,Walmsley}. These effects, in particular, the
Hong-Ou-Mandel dip \cite{dip}, were however obtained through a
quantum measurement {\it on a biphoton} alone. In the setup
considered in this paper the goal is to match one of the photons in a
pair to an $external$ optical field. Hence the difference in
requirements.

 To make the calculation of the purity
parameter more practical we rewrite Eq. (\ref{P_temp}) in terms of
the experimentally accessible full temporal width at half intensity
maximum (FWHM) of the pump pulse $\tau_p=2\sqrt{2\ln2}/\sigma_p$ and
the spectral FWHM of the spectral filter transmission function
$w_t=2\sqrt{\ln2}\sigma_t$. Approximating Eq. (\ref{P_temp}) for
$\mu_t\ll 1$, we obtain
\begin{equation}  \label{P_temp_simp}
P_{\rm temp}\approx 1-\mu_t^2=1-\frac{w_t^2\tau_p^2}{32(\ln2)^2}.
\end{equation}

Our next goal is to determine and optimize the fidelity of mode
matching between the DFG and CPP modes. Substituting the correlation
function of the coherent alignment field
$\Gamma_A(\kvec,\omega,\kvec',\omega')=
E_{A}^*(\kvec,\omega)E_{A}(\kvec',\omega')$ into Eq.
(\ref{EqnKlass}), we find for the DFG field:
\begin{equation}
  \Gamma_{\rm DFG}(\omega,\omega') = \Gamma_0^{\rm DFG}
  \exp\biggl(-\frac{(\omega-\omega_s^0)^2 +
  (\omega'-\omega_s^0)^2}{( \sigma_p^2 + \sigma_A^2)} \biggl). \nonumber
\end{equation}

With this, using Eqs.\,(\ref{GammaPhi}) and (\ref{EqnDefMM}) we
obtain the mode matching factor:
\begin{equation} \label{MMGauss}
  M (\mu_{t},\mu_{A})=  \left(\sqrt{\frac{1+\mu_{A}^2}{(1+\mu_{A}^2/2+\mu_{t}^2)(1+\mu_{A}^2/2)}}\right),
\end{equation}
where $\mu_{A} = \sigma_{A}/\sigma_p$. For a given $\mu_t$, $M
(\mu_{t},\mu_{A})$ reaches its maximum at $\mu_A^{\rm
max}=\sqrt{\sqrt{1+2\mu_t^2}-1}$, which can be approximated as
$\mu_A\approx\mu_t$ for small values of $\mu_t$. {\it The DFG field
models the CPP mode optimally when the width of the alignment pulse
in the frequency-momentum space is equal to that of the transmission
function of the filter}.



In Fig. \ref{AbbMM}(b) we plot $M(\mu_{t},\mu_{A}^{\rm max})$ as a
function of $\mu_{t}$. We see that the mode matching parameter
approaches its theoretical limit $\sqrt{P_{\rm temp}}$ at low values
of $\mu_{t}$; in fact, the difference does not exceed 0.5\% for
$\sigma_t<\sigma_p/2$. This shows that the presented technique of
modeling the CPP mode with a classical wave is indeed effective as
long as the filtering in the trigger channel is sufficiently tight.

Note that in the limit of narrowband filtering the parameters of the
alignment beam do not play an important role. Curve (c) in
Fig.\,\ref{AbbMM} shows the behavior or $M(\mu_{t},\mu_{A})$ with
$\mu_A\equiv 0$. Instead of an alignment field of optimal parameters,
simply a plane wave is used. Although the level of mode overlap is
not as high as for the optimal case, the difference is negligible for
low $\mu_t$.

\subsection{Time-space representation}
To gain a further insight into the phenomena considered in this paper
we calculate the CPP mode purity parameter in the space-time
representation rather than the frequency-momentum representation we
used so far. In this subsection we work with correlation functions
defined as $\tilde\Gamma(\rvec,t,\rvec',t')=\left<\tilde
E^{(-)}(\rvec,t)\,\tilde E^{(+)}(\rvec',t')\right>$, with the mode
matching and purity parameters redefined analogously.

We employ the same assumptions as in the previous subsection but do
the calculation for the spatial domain. To facilitate visualizing the
physics involved, we utilize Klyshko's advanced wave model and
perform the entire calculation classically.

The transverse correlation function of the DFG mode is obtained from
Eq. (\ref{Eproduct}) and is as follows:
\begin{equation}\label{Gamma_s_r}
\tilde\Gamma_s(\rvec,\rvec')=|\beta|^2\tilde E_p^{(-)}(\rvec)\tilde
E_p^{(+)}(\rvec')\tilde\Gamma_t^*(\rvec,\rvec'),
\end{equation}
where $\rvec$ denotes the transverse radius vectors in the crystal
plane and $\tilde\Gamma_t(\rvec,\rvec')$ is the correlation function
of the advanced wave in the plane of the crystal. In writing this
equation we made use of the fact that the pump pulse is a coherent
wave.

The photon counter behind the spectral filter pinhole is replaced,
according to the Klyshko model, by a source generating spatially
incoherent light backwards in space and time. The light emitted by
the source passes through the pinhole and is collimated by the
focusing lens (Fig.\,\ref{PDC2}(b)). The correlation function of the
advanced wave in the plane of the nonlinear crystal is then equal to
that in the plane of the focusing lens. The latter is determined in
the far-field approximation using the Van Cittert-Zernike theorem
\cite{BW}:

\begin{equation} \label{Gamma_t_r}
\tilde\Gamma_t(\rvec,\rvec')=\int T(\Rvec) e^{-i (k_t/F)\Rvec\cdot
(\rvec-\rvec')} d^3R ,
\end{equation}
where $k_t=2\pi/\lambda_t$ is the trigger wavenumber, $F$ is the
focal length of the lens and $T(\Rvec)$ is the pinhole transmission.

\subsubsection{Gaussian filter}
For a Gaussian filter (\ref{T_omega}), replacing
$\kvec_{\bot}=k_t\Rvec/F$ and performing a Fourier transform
according to Eq. (\ref {Gamma_t_r}) we obtain the correlation
function of the advanced wave:
\begin{equation} \label{Gamma_t_R_G}
  \tilde\Gamma_t(\rvec,\rvec')=\tilde\Gamma_t^0 \exp\left(-(\kappa_t|\rvec-\rvec'|/2)^2\right).
\end{equation}
Substituting it into Eq.(\ref{Gamma_s_r}) and writing for the pump
 field (\ref{EpA_omega}) $\tilde
E_p^{(+)}(\rvec)\propto\exp(-(\kappa_p|\rvec|/2)^2)$, we determine
the spatial correlation function of the signal mode and its purity
parameter:
\begin{equation} \label{P_sp_G}
P_{\rm sp}=\frac{\int d^3r\,  d^3r' \,
  \tilde\Gamma_s(\rvec,\rvec')  \tilde\Gamma_s^*(\rvec,\rvec')}
  {(\int d^3r  \,
  \tilde\Gamma_s(\rvec,\rvec))^2}=\frac{1}{1+2\kappa_t^2/\kappa_p^2}.
\end{equation}
This expression is clearly analogous to Eq. (\ref{P_temp}). The
absence of a square root is explained by the two-dimensional
character of spatial mode matching.

The light emitted by Klyshko's virtual source is completely
incoherent. However, {\it as the advanced wave passes through a
narrow aperture, it gains some degree of transverse coherence
according to the Van Cittert-Zernike theorem. Because the nonlinear
interaction is restricted to the area where the pump field is
present, the resulting signal (DFG) field is also partially coherent
provided the pump beam diameter is smaller than the transverse
coherence length of the advanced wave.} This explains why the
advanced wave, in spite of its own incoherence, may generate a highly
coherent CPP signal.

\subsubsection{Cylindrical filter}To make the above calculation more
useful for practical applications, consider a spatial filter not of
Gaussian shape, but of top-hat shape, i.e. the pinhole transmits all
the light within its radius $\rho$. In this case, the correlation
function of the advanced wave, calculated using Eq.
(\ref{Gamma_t_r}), is given by
\begin{equation}  \label{Gamma_t_R_C}
\Gamma_t(\rvec,\rvec')=\frac{2J_1(k_t\rho|\rvec-\rvec'|/F)}{(k_t\rho|\rvec-\rvec'|/F)},
\end{equation}
where $J_1(x)$ is the first order Bessel function. Approximating
$2J_1(x)/x\approx 1-x^2/8$ for small $x$ and comparing the
correlation functions (\ref{Gamma_t_R_G}) and (\ref{Gamma_t_R_C}) we
find that the two functions behave similarly on small spatial scales
if $\kappa_t=k_t\rho/(F\sqrt{2})$. Substituting the latter identity
into Eq. (\ref{P_sp_G}) and expressing $\kappa_p$ through the FWHM
diameter $d_p$ of the pump beam ($\kappa_p=2\sqrt{2\ln2}/d_p$) we
find in the limit of tight filtering
\begin{equation}  \label{P_sp_C}
P_{\rm sp}\approx 1-\left(\frac{\pi\rho d_p}{\sqrt{2\ln2}\lambda_t
F}\right)^2,
\end{equation}
where $\lambda_t$ is trigger wavelength. Equations
(\ref{P_temp_simp}) and (\ref{P_sp_C}) provide the means for
evaluating the CPP mode purity factor from a set of parameters that
are readily measurable in an experiment. The mode matching efficiency
is then evaluated using inequality (\ref{limit}).


\section{Experimental considerations}
As we have shown in the previous section, tight filtering in the
trigger channel, both spatial and temporal, is the key to obtaining a
CPP ensemble which approaches a pure state and can be coupled into a
classical optical mode. In experimental practice, reducing the width
of spatial and spectral filters lowers the trigger count rate and
increases the relative fraction of dark counts. As seen from Eq.
(\ref{P_temp_simp}), a reasonable compromise (dependent on a
particular application) is a spectral filter FWHM on the order of the
inverse duration of the pump pulse. A favorable size for the pinhole
in the spatial filter is obtained from Eq. (\ref{P_sp_C}) and should
be such that the imaginary coherent wave propagating backwards
through the pinhole would create a diffraction spot in the optical
plane of the crystal which is several times larger than the diameter
of the pump beam.

An important experimental limitation is imposed by the existing
technology of making optical coatings. Interference bandpass filters
narrower than 1 $\rm\AA$  are not available or very expensive. {\it
Ultrashort (less than a few picoseconds) laser pulses must therefore
be used} to pump the downconverter so that the trigger filter
bandwidth can be made sufficiently narrow in comparison with the pump
linewidth.

The freedom of choice of the optimally matching classical pulse is
also limited. While its spatial parameters can be varied in a wide
range, its temporal width is set by the master laser and cannot be
changed easily. Using laser pulses of non-optimal width results in a
reduction of the temporal mode matching efficiency. If the width of
the matching Gaussian pulse differs from that of the CPP mode by a
factor of $\alpha$, the mode matching is reduced by a factor of
\begin{equation}
f(\alpha)=\frac{2\alpha}{\alpha^2+1} .
\end{equation}

\begin{figure}
\begin{center}
\includegraphics[width=0.45\textwidth]{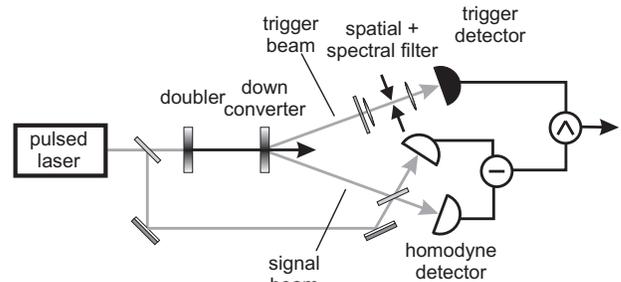}
\caption{ \label{FockFig} Scheme of the experiment on quantum
tomography of single-photon states}
\end{center}
\end{figure}

As a practical example of solving the mode matching problem we
consider the experiment of Lvovsky {\it et al.} \cite{Fock} on
homodyne tomography of the single-photon Fock state
(Fig.\,\ref{FockFig}). In this experiment, $\lambda=790$ nm,
$\sqrt{2}\tau_p=$1.6 ps\footnote{A factor of $\sqrt{2}$ appears due
to the frequency doubling} master laser pulses were frequency doubled
and then down-converted in a type-I frequency degenerate
configuration. Such a scheme permitted to use a fraction of the
original laser radiation as the local oscillator for the balanced
homodyne detector. The photons emitted into the trigger channel were
filtered by a combination of a $0.4$-nm ($w_t=1.2\times10^{12}\
\rm{s}^{-1}$) FWHM interference filter and a spatial filter
consisting of a $F=80$ mm focal length lens and a $2\rho=50$ $\mu$m
pinhole. The FWHM diameter of the pump beam was $d_p=0.34$ mm. The
laser frequency was centered at the transmission peak of the spectral
filter which insured the coincidence of the center frequencies of the
CPP and the local oscillator (LO).

In order to perform an efficient homodyne measurement it was
necessary to achieve good mode overlap between the local oscillator
and the CPP. Since at the time of the measurement no alignment beam
could have been present in the trigger channel, the procedure of
preparing the matching classical mode described in Section
\ref{SecModel} was applied in two steps. In the first step, a
fraction of the master laser field was inserted into the trigger
channel and synchronized with the pump pulses. This field generated
DFG emission through its nonlinear interaction with the pump and thus
fulfilled the function of the alignment beam. The LO beam was
overlapped with the DFG beam on a 50-\% beamsplitter and interference
between the two classical fields was observed in one of the
beamsplitter output ports. The visibility of the interference pattern
was maximized by varying the spatial parameters of the LO beam with a
3-lens telescope and steering mirrors.  In the second step, the
alignment beam was blocked and a tomography measurement was performed
using the same beamsplitter for balanced homodyning.

The task of evaluating the mode matching between the LO and CPP modes
thus splits into two parts. First, the purity parameter $P$ of the
CPP mode needs to be evaluated $theoretically$ to establish the upper
limit for the mode matching between the CPP and DFG waves. Second,
the level of mode matching $M_{\rm cl}$ between the classical DFG and
LO waves has to be determined $experimentally$ from the visibility of
the interference pattern. The overlap between the CPP and LO modes
can then be evaluated as a product $M=M_{\rm cl}\sqrt{P}$.

Although the down-conversion occurred in a non-collinear
configuration, the angle between the down-conversion channels and the
pump beam was relatively small ($6.8^\circ$) so the approximations
outlined in the beginning of Section \ref{SecCalc} were applicable to
the system. The signal beam walk-off was eliminated by using the
``hot spot" configuration of the down-converter \cite{HotSpot} and
the group velocity dispersion effects were negligible
\cite{HHthesis}. Applying Eqs. (\ref{P_temp_simp}) and (\ref{P_sp_C})
to the actual experimental parameters we find the values of $P_{\rm
temp}=0.85$ and $P_{\rm sp}=0.87$ for the temporal and spatial purity
parameters of the CPP mode, respectively. This corresponds to a
maximum achievable mode matching efficiency of $\sqrt{P}=\sqrt{P_{\rm
temp}P_{\rm sp}}=0.86$.

The maximum visibility of the interference fringes observed between
the DFG and LO waves was equal to 0.83 which corresponds to a mode
matching factor $M_{\rm exp}=0.69$. In order to obtain $M_{\rm cl}$,
this value needs to be corrected to accommodate for the temporal
properties of the alignment pulse. While its $spatial$
characteristics were optimized according to the requirements
established in Section \ref{SecCalc} (the alignment beam was made
broad and collimated so it passed well through the spatial filter),
its $temporal$ properties were beyond our control. The alignment
field was not narrowband (as required), which resulted in a different
linewidth of the DFG field as compared to the CPP mode. The nonlinear
interaction between the second harmonic (pump) and the fundamental
(alignment) waves produces a DFG wave whose linewidth is broader by a
factor $\sqrt{3}$ than the fundamental. On the other hand, the
spectral linewidth of the CPP mode in the limit of narrow filtering
would mimic that of the pump, which is $\sqrt{2}$ times the
fundamental \cite{Grangier}. If a narrowband alignment beam were
available, the mode overlap between the LO and DFG waves would have
been by a factor of $f(\sqrt{2})/f(\sqrt{3})=1.09$ higher than the
one actually observed. We find $M_{\rm cl}=M_{\rm
exp}f(\sqrt{2})/f(\sqrt{3})=0.75$.

We calculate the overall factor of spatiotemporal mode matching
between the LO and CPP waves as $M=\sqrt{P}\times M_{\rm cl}=0.65$.
This number is in agreement with the value of $0.69\times0.95$ quoted
in Ref. \cite{Fock}.




\section{Conclusion}
We have investigated the spatiotemporal optical mode of the
single-photon Fock state prepared by conditional measurements on a
biphoton born in the process of parametric down-conversion and the
possibilities of matching it with a classical wave. Our theory,
developed using the density-matrix formalism, shows that in order to
obtain a pure single-photon state in the signal channel it is
essential to provide narrow spatiotemporal filtering in the trigger
channel. Only in this case can efficient mode matching be achieved.
The theoretical limit of mode matching can be expressed in terms of
the CPP mode purity factor which is readily determined as a function
of the experimental parameters.

We have shown that the optical mode of the CPP is identical to that
of a classical wave generated due to a nonlinear interaction of the
pump wave and Klyshko's advanced wave. Based on this knowledge we
proposed and implemented an experimental method of modeling the CPP
mode by using a narrowband alignment beam in place of the advanced
wave. The difference frequency field generated in such an arrangement
matches the CPP mode with an efficiency that approaches the
theoretical limit.



Finally, we have discussed how the mode matching efficiency can be
evaluated and optimized in a practical experimental arrangement.

This project is sponsored by the Deutsche Forschungsgemeinschaft. A.
L. is supported by the Alexander von Humboldt foundation. We thank
Prof. J. Mlynek and Dr. H. Hansen for helpful discussions.


\end{document}